\documentclass{INTERSPEECH2023}
\usepackage{caption}
\usepackage{subcaption}
\usepackage{multirow}
\usepackage{hyperref}
\usepackage{tablefootnote}
\usepackage{comment}
\usepackage{cite}
\usepackage{adjustbox}
\hypersetup{colorlinks=false,breaklinks=true}
\DeclareMathOperator{\plog}{plog}
\DeclareMathOperator{\arcsinh}{arcsinh}

\interspeechcameraready

\title{mdctGAN: Taming transformer-based GAN for speech super-resolution with Modified DCT spectra}
\name{Chenhao Shuai$^{1,3,4}$\sthanks{\quad Major work done during undergraduate at CQUPT \& Brunel}\ \ , Chaohua Shi$^{2,3,4*}$, Lu Gan$^3$, Hongqing Liu$^4$\sthanks{\quad Intelligent Speech and Audio Lab, CQUPT}}
\address{
  $^1$Nanyang Technological University, Singapore \quad
  $^2$Xidian University, China \\
  $^3$Brunel University London, UK \quad
  $^4$Chongqing University of Posts and Telecommunications, China}
\email{shua0003@e.ntu.edu.sg, chshi2004@gmail.com, lu.gan@brunel.ac.uk, hongqingliu@cqupt.edu.cn}

\begin{document}

\maketitle
 
\begin{abstract}
Speech super-resolution (SSR) aims to recover a high resolution (HR) speech from its corresponding low resolution (LR) counterpart. Recent SSR methods focus more on the reconstruction of the magnitude spectrogram, ignoring the importance of phase reconstruction, thereby limiting the recovery quality. To address this issue, we propose \textit{mdctGAN}, a novel SSR framework based on modified discrete cosine transform (MDCT). By adversarial learning in the MDCT domain, our method reconstructs HR speeches in a phase-aware manner without vocoders or additional post-processing. Furthermore, by learning frequency consistent features with self-attentive mechanism, \textit{mdctGAN} guarantees a high quality speech reconstruction. For VCTK corpus dataset, the experiment results show that our model produces natural auditory quality with high MOS and PESQ scores. It also achieves the state-of-the-art log-spectral-distance (LSD) performance on 48 kHz target resolution from various input rates. Code is available from \url{https://github.com/neoncloud/mdctGAN}
\end{abstract}

\section{Introduction}

Speech super-resolution (SSR) is the task of recovering high-resolution (HR) speech from low-resolution (LR) speech. SSR presents many practical applications in fields such as teleconferencing and speech recognition. It is crucial to correctly reconstruct the phase information when recovering the high frequency components. Recent frequency domain-based SSR research \cite{MelGAN,WaveNet,HiFiGAN,nvsr,DCCRN, DCCRN+, WSRGlow,Nonparallel} has primarily focused on recovering amplitude information, and it often requires additional steps to reconstruct the missing phase.

% by using Griffin-Lim algorithm~\cite{GF} or vocoders to 

%Vocoders\cite{MelGAN,WaveNet,HiFiGAN} are popular choices to reconstruct the raw waveform, which recover the phase by discriminating on time-domain. 
Vocoders~\cite{MelGAN,WaveNet,HiFiGAN}, which recover the phase by discriminating in the time domain, are frequently used to reconstruct the raw waveform.
Utilizing vocoders, a two-stage speech super-resolution method~\cite{nvsr} was proposed to predict high-resolution speech mel-spectrograms at first, then applying a vocoder for raw waveform reconstruction. Though inspiring, such a two-stage generation process can lead to instability during training. Alternatively, recent approaches attempt to model phase and magnitude (or real and imaginary part of complex-valued spectrograms) by complex-valued neural networks or fusion modules~\cite{DCCRN, DCCRN+, WSRGlow}. The waveform can then be reconstructed by simply applying an inversion. However, the convergence of complex neural networks is not guaranteed. Separating the treatment of the complex-valued features leads to implicit modeling of phase and magnitude. 

To address these issues, we propose to perform SSR in the modified discrete cosine transform (MDCT) domain, a real-valued, lossless transform, which enables a joint magnitude and phase estimation. To that aim, we propose \textit{mdctGAN}, a frequency-attentive, phase-aware SSR network with a transformer-based~\cite{Vaswani} conditional Generative Adversarial Networks (cGAN)~\cite{GAN, pix2pixHD} architecture. In the network design, we introduce a transformer bottleneck stack in the generator network for global attention on frequency-consistent features. Transformer-based models are inherently data-hungry. However, the current VCTK dataset~\cite{WaveNet} used for the SSR task is not sufficient to fully exploit the potential of our model. Hence, we added the HiFi-TTS dataset~\cite{HiFi-TTS_data}, a large-scale, high-quality speech synthesis corpus, for pre-training. And we fine-tuned our model on the VCTK corpus~\cite{WaveNet} to achieve an output sampling rate up to 48 kHz. 
Specifically, the contributions of this work are as follows.
\begin{itemize}
    \item We propose a vocoder-free method that performs speech super-resolution with transformer-based cGAN. It outperforms previous works in LSD scores. Meanwhile, it also achieves a high score in subjective tests.
    
    \item %We implement a differentiable, fast MDCT and its inverse transform layer in PyTorch. 
    We develop the pseudo-log operation for dynamic compression of MDCT coefficients, which is essential to the production of phase-aware, high-quality speech.
    
    \item We show by experiments that our models can learn and thereby generate the phase information encoded in the MDCT coefficients well, demonstrating the great potential in producing high-quality speech audio.
\end{itemize}

\section{Proposed Method}
\subsection{The MDCT-based processing}
\subsubsection{Basics of the MDCT}
The reconstruction of the speech phase is crucial to the quality of the generated speech. The commonly used mel-spectrograms do not contain the phase information of the audio signal and therefore require additional algorithms for phase recovery. For this reason, we propose SSR in the MDCT domain, which guarantees the phase-aware speech reconstruction with a real-valued spectrogram. 
MDCT is widely used in audio compression, e.g. mp3, ac3~\cite{mp3}. Just as the short-time Fourier transform (STFT), the audio is first split into blocks, with a 50\% overlap between each block. Then, a forward MDCT is applied to each block, given below:
%\small
%\small
\begin{equation}
\begin{aligned}
%\Scale[0.5]{
&X_i[k] = \sum_{n=0}^{N-1} w[n]x_i[n]\cos\left(\frac{2\pi}{N}(n+n_0)(k+\frac{1}{2})\right)\\
    &k = 0,...,N/2-1,%}
\end{aligned}
\end{equation}
%\normalsize
%\normalsize
where $n_0 = N/4+1/2$, $x_i[n]$ is the $n$-th sample in the $i$-th block of the audio, and $w[n]$ is the window function applied to each block to reduce spectrum leakage. The corresponding inverse MDCT (iMDCT) of the $i$-th block is:
%\small
\begin{equation}
\begin{aligned}
    &x'_i[n] = \frac{4}{N}w[n]\sum_{k=0}^{N/2-1} X_i[k]\cos\left(\frac{2\pi}{N}(n+n_0)(k+\frac{1}{2})\right)\\
    &n = 0,...,N-1.
\end{aligned}
\end{equation}
%\normalsize
To produce the full audio, each recovered block should be overlap-added together to eliminate the aliasing. 
In this paper, $w[n]$ was chosen as the Kaiser-Bessel-derived window~\cite{mp3}. And we implemented fast MDCT/iMDCT modules using FFT.

\begin{figure}[t]
\centering
  \includegraphics[width=0.94\columnwidth]{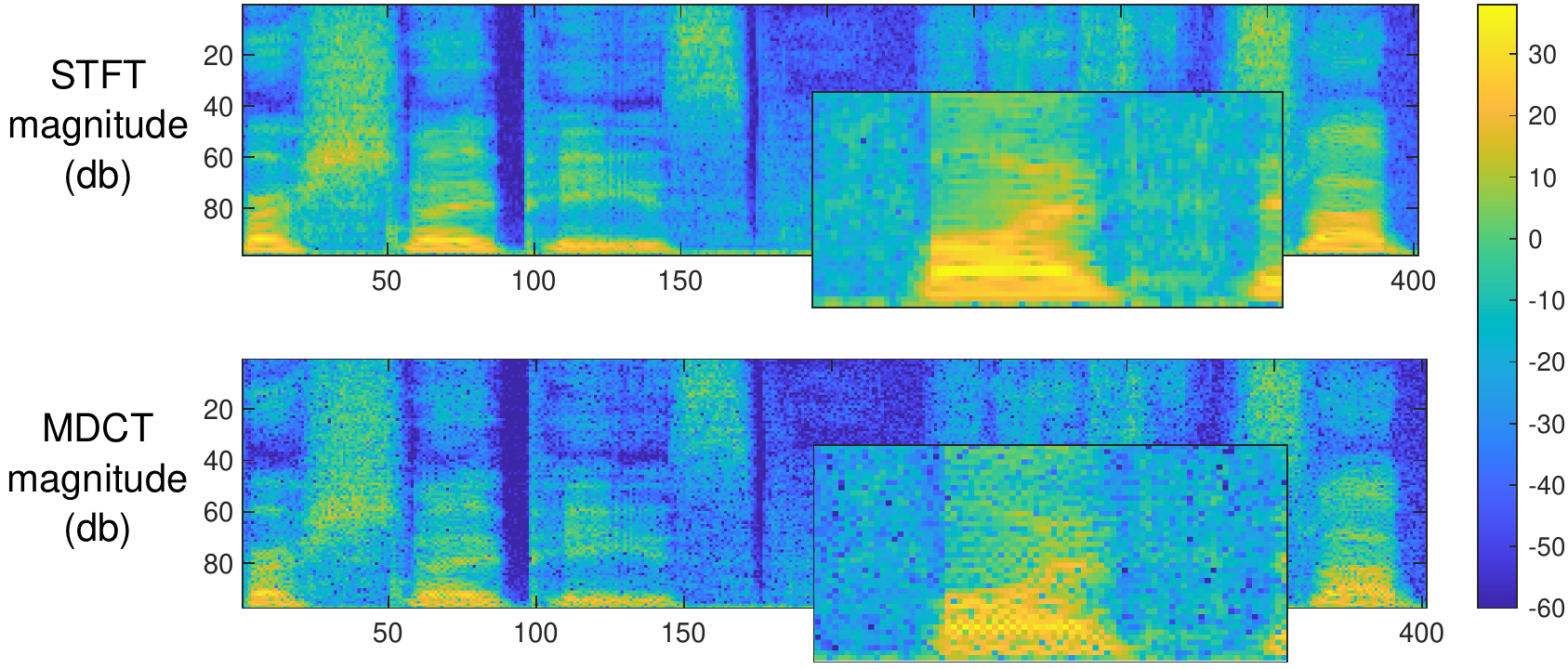}
  \caption{STFT magnitude vs MDCT magnitude in decibel scale, sampled from VCTK corpus s225-003. Just as the STFT spectrogram, the MDCT spectrogram also reveal rich acoustic features, such as resonant peaks in speech.}
  \label{fig:mdct}
\end{figure}

This time-frequency domain representation is similar to that of the STFT, as shown in Figure~\ref{fig:mdct}, where the resonant peaks of the speech signal can be observed, indicating that it is also an effective representation of the frequency component distribution of the signal. The MDCT has the following advantages: 
\begin{itemize}
    \item It is a real-valued, invertible linear transform 
    so that the output is well compatible with existing deep learning neural networks.
    \item The phase information of raw waveform is encoded into the \textit{sign} of MDCT coefficient, allowing the neural networks to model and reconstruct phase and amplitude of signal components with MDCT spectra only.
\end{itemize}
% \begin{comment}
% \subsubsection{Fast MDCT implementation}
% With the native FFT API provided by PyTorch, we have implemented a differentiable, fast MDCT/iMDCT that efficiently converts speech to spectrum as follows: 
% \small
% \begin{align}
%     &X_i[k] = \mathcal{R}e\left\{R_{post}^{k}FFT_{N}\left[w[n]x_i[n]R_{pre}^{n}\right] \right\}\notag\\
%     &k =\ 0,\ \dots,\ N/2-1.
% \end{align}
% \normalsize
% Here, pre-twiddling $R_{pre}^{n} = e^{-j\frac{\pi}{N}}$ and post-twiddling $R_{post}^{k}=e^{-j\frac{2\pi}{N}n_0(k+\frac{1}{2})}$ is needed for every block. 
% % It can be accelerated on GPU by broadcasting mechanism in PyTorch. 
% Similarly, we have the fast iMDCT:
% \small
% \begin{align}
%     &x_i'[n] = \frac{4}{N}w[n]\mathcal{R}e\left\{R_{pre}^{n}FFT_{N}\left[X_i[k]R_{post}^{k}\right]\right\}\notag\\
%     &n = 0,\ \dots,\ N-1.
% \end{align}
% \normalsize
% \end{comment}
\subsubsection{Pseudo-log dynamic range compression}
\begin{figure}[t]
    \centering
    \includegraphics[width=0.94\columnwidth]{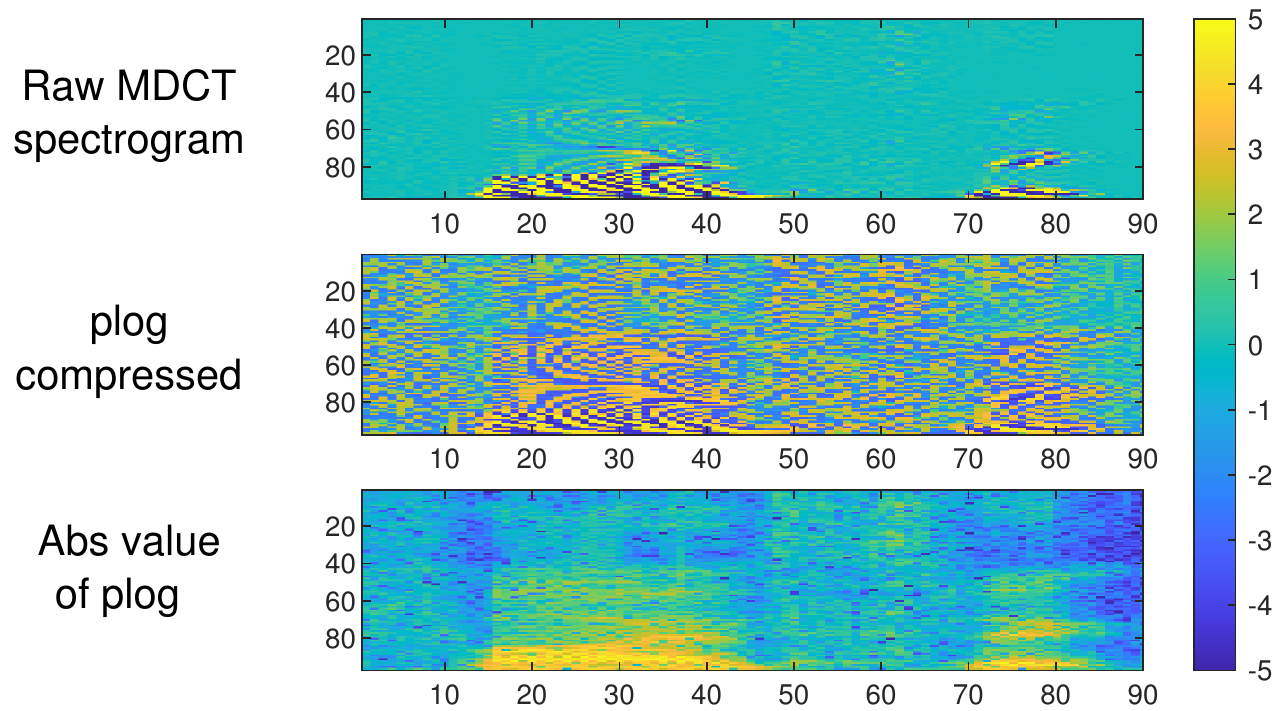}
    \caption{Dynamic range compression with pseudo-log. Note that the MDCT coefficients after pseudo-log compression exhibits similar patterns to decibel-scaled STFT magnitude.}
    \vspace{-0.2cm}
    \label{fig:plog_vs_raw}
\end{figure}
In a typical speech processing pipeline, small frequency components are revealed using decibel-scaled STFT magnitudes rather than the raw spectrogram. However, MDCT encodes the phase information into the sign of the coefficients, and unfortunately, we cannot perform logarithmic on negative values. In order to maintain the sign of the coefficients while compressing the dynamic range of the signal, we introduce a \textbf{pseudo-logarithmic} operation based on
%In this subsection, we describe how to compress the dynamic range of raw MDCT coefficients based on 
$\arcsinh(x)$:
%\small
\begin{equation}\label{eq:plog}
    \plog(x) = \frac{\arcsinh(x)}{\ln(10)} = \log_{10}(x+\sqrt{x^2+1}).
\end{equation}
%\normalsize

We illustrate the impact of dynamic compression in Figure~\ref{fig:plog_vs_raw}. The ablation experiments demonstrated that without the $\plog$ operation, the model cannot be trained. The $\plog$ function in (\ref{eq:plog}) has the following desirable properties.
\begin{itemize}
\item It is differentiable with respect to $\mathbb{R}$ and is oddly symmetric. Hence, it preserves the polarity of both positive and negative MDCT coefficients.
\item By using the asymptotic expansion of $\arcsinh(x)$, it can be shown that as $x\rightarrow+\infty$, $\plog(x)\rightarrow\log_{10}(2x)$. It compresses the dynamic range similar to decibel-scale of MDCT (or STFT) magnitude, as shown in Figure~\ref{fig:plog_vs_raw}.
\end{itemize}

To make better use of the non-linear interval of the $\plog$ function, we multiply the raw MDCT coefficients with a gain of $\alpha=10^3$. We have found that the histogram distributions of the transformed spectra mostly lie in the interval of $[-5,5]$, so we scale them to $[-1,1]$ by dividing 5. Alternating positive and negative patterns are frequently observed in an MDCT spectrogram, shown in Figure~\ref{fig:plog_vs_raw}. To prompt the model with magnitude information, we also appended the corresponding absolute values to the input, which also normalised to the $[-1,1]$ interval, guiding the network to produce clearer high-frequency details.

\subsection{Network architecture}
\begin{figure*}[t]
\centering
  \includegraphics[width=\textwidth]{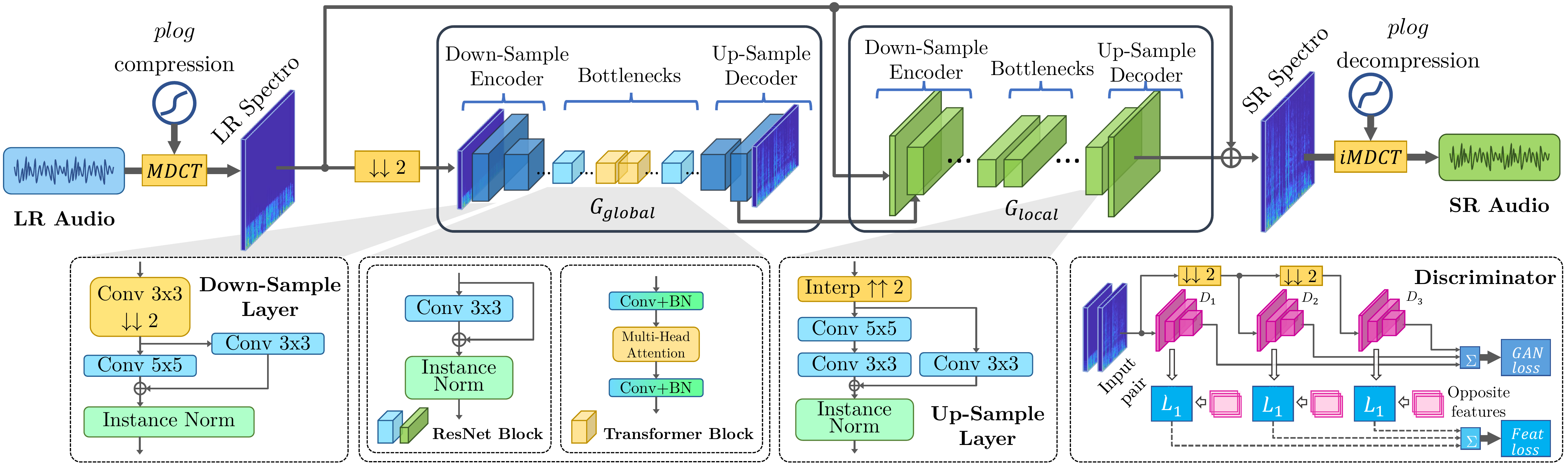}
  \caption{Architecture of the proposed network. Spectrogram generation is from coarse to fine, $G_{global}$ is responsible for modelling global, large scale features.  reduce the computational burden, $G_{global}$ runs on 2x downsampled inputs. To enrich the detail of the generated spectrogram, $G_{local}$ will fuse the features from $G_{global}$ and the original input to reconstruct the detailed high frequency components. The architecture of the proposed multi-scale spectrogram discriminator is shown at the bottom right corner.}
  \label{fig:G_net}
\end{figure*}
In this work, we mainly focus on training a network $G$ to map the $LR$ spectrogram to $\hat{\bm{SR}}'=G(\bm{LR})$ by minimizing the error between $\hat{\bm{SR}}'$ and $\bm{HR-LR}$. The final output is defined as $\hat{\bm{SR}} = \hat{\bm{SR}}'+\bm{LR}$. Figure~\ref{fig:G_net} shows the overview of the proposed mdctGAN architecture. We want our model to generate the fine structure of the spectrogram (e.g. the resonant peaks of speech) while remaining globally consistent with the base frequency component. Inspired by work on image translation~\cite{pix2pixHD}, we have designed a transformer-based generator architecture that works from coarse to fine, as well as a discriminator that judges the spectrogram from multiple scales.

\subsubsection{Generator}

As shown in Figure \ref{fig:G_net}, our generator consists of two sub-networks: $G_{global}$ (the global generator network, working on the $\downarrow\downarrow 2$ input) and $G_{local}$ (the local enhancer network, working on the full size spectrogram) achieving coarse-to-fine spectrogram generation~\cite{pix2pixHD}. Both sub-networks use Unet-like architectures, yet differ in size and depth, consisting of three sub-modules: a spectrogram encoder that extracts features at multiple scales through cascaded convolution layers, a bottleneck block stack and a decoder that progressively up-samples features with bilinear interpolation (denoted by Interp $\uparrow\uparrow 2$ in the Up-Sample Layer). The bottleneck of $G_{global}$ also contains Transformer blocks (yellow boxes in the middle of $G_{local}$) for learning frequency consistent features. Compared with the original $stride=2$ transpose-convolution used in~\cite{pix2pixHD}, this upsampling operator can reduce the checkerboard artefacts more effectively, as demonstrated in the ablation study of Section~\ref{sec:ablation}. Finally, the model produces a full-band SR spectrogram by summing the residual path of the input LR spectrogram. %The overall output is the sum of the low frequency part of the input and the generated high frequency part.

\subsubsection{Discriminator}
To achieve our vocoder-free SSR goal, we only use a multi-scale discriminator supervised in the MDCT domain. It contains a total of 3 discriminators, and all have the same network structure but operating at 3 different scales by downsampling the input. All discriminators follow the Patch-GAN's architecture, with basic blocks of cascaded \texttt{Convolution}-\texttt{InstanceNorm}-\texttt{LeakyReLU} ($slope = 0.2$) layers. This design 
allows the generator to efficiently produce both globally consistent spectrograms (coarse-level supervision) and finer detail information (fine-level supervision).

\subsection{Loss function}

In this work, the total loss function $L_t$ consists of an adversarial loss and a feature matching loss, which is similar to that in~\cite{pix2pixHD}. %Specifically, the total loss function $L_t$ is given as
\small
\begin{equation}
\begin{split}
L_t=\min_{G}&\left[\left(\max_{D_1,D_2,D_3}\sum_{i=1}^3 V_{GAN}(G,D_i))\right)\right.\notag\\
    &+\left.\lambda_{feat}\sum_{i=1}^3 V_{Feat}(G,D_i)\right],
\end{split}
\end{equation}
\normalsize
where $\lambda_{feat}$ is the gain of feature matching loss. Here,  $V_{GAN}(G,D_i)$ represents the adversarial loss of each $D_i$
%\small
\begin{equation}\nonumber%\label{eq:gan_ssr}
\begin{split}
V_{GAN}(G,D_i)=&\mathbb{E}_{(\bm{LR},\bm{HR})\sim p_{\text{data}}(\bm{LR},\bm{HR})}[\log(D_i(\bm{LR},\bm{HR}))]\\
    +&\mathbb{E}_{\bm{LR}\sim p_{\text{data}}(\bm{LR})}[\log(1-D_i(G(\bm{LR})))].
    \end{split}
\end{equation}
%\normalsize
As shown in the bottom right corner of Figure \ref{fig:G_net}, $\{D_i\}_{i=1}^3$ process the original signal, down-sampled versions with decimation factors of 2 and 4, respectively. In this way, the discriminator network processes inputs from coarse to fine. Each feature loss function $V_{Feat}(G,D_i)$ takes the following form
\small
\begin{equation}
\begin{split}
    &V_{Feat}(G,D_i) \notag\\
    =&\mathbb{E}_{(\bm{LR},\bm{HR})\sim p_{\text{data}}(\bm{LR},\bm{HR})}\frac{1}{N_k}\sum_{i=1}^{3}\sum_{k}\left\lVert\bm{F}^{ik}_{pos}-\bm{F}^{ik}_{neg}\right\rVert_{1},
    \end{split}
\end{equation}
\normalsize
where $\bm{F}_{pos}^{ik}=D_i^k(\bm{LR},\bm{HR})$ corresponds to intermediate feature maps at the $k$-th layer of the $i$-th discriminator $D_i$ for the pair $(\bm{LR},\bm{HR})$. Likewise, $\bm{F}^{ik}_{neg}=D_i^k(\bm{LR}; G(\bm{LR}))$ represents that of $(\bm{LR} , G(\bm{LR}))$, denoted as ``Opposite Features" in Figure \ref{fig:G_net}. 
As MDCT encodes the phase information, we do not need to design an additional time domain penalty term.

\section{Experiments}
\begin{table*}[!h]
\begin{center}
\begin{minipage}{0.55\textwidth}
    \centering
    \caption{A comparison of SNR and LSD scores with 48 kHz target.}
    \renewcommand{\arraystretch}{1.2}
    \resizebox{!}{0.063\textheight}
    {%
    \begin{tabular}{cccccccccc}
    \hline
    \multirow{3}{*}{\begin{tabular}[c]{@{}c@{}}\\ Model\\ Name\end{tabular}} & \multirow{3}{*}{\begin{tabular}[c]{@{}c@{}}\\ \# Params\\ \end{tabular}} & \multicolumn{8}{c}{Input sampling rate} \\ \cline{3-10} 
     &  & \multicolumn{2}{c}{\begin{tabular}[c]{@{}c@{}}24 kHz\\ ($2\times$ SR)\end{tabular}} & \multicolumn{2}{c}{\begin{tabular}[c]{@{}c@{}}16 kHz\\ ($3\times$ SR)\end{tabular}} & \multicolumn{2}{c}{\begin{tabular}[c]{@{}c@{}}12 kHz\\ ($4\times$ SR)\end{tabular}} & \multicolumn{2}{c}{\begin{tabular}[c]{@{}c@{}}8 kHz\\ ($6\times$ SR)\end{tabular}} \\ \cline{3-10} 
     &  & $SNR\uparrow$ & $LSD\downarrow$ & $SNR\uparrow$ & $LSD\downarrow$ & $SNR\uparrow$ & $LSD\downarrow$ & $SNR\uparrow$ & $LSD\downarrow$ \\ \hline \hline
    AudioUNet & 70.9M & 22.68 & 1.01 & - & - & 17.15 & 2.24 & - & - \\
    MUGAN & 70.9M & 24.81 & 0.90 & - & - & 16.87 & 2.12 & - & - \\
    WSRGlow\tablefootnote{Reported by NU-Wave 2} & 229M & 26.60 & 0.70 & 22.60 & 0.84 & 21.20 & 0.94 & 18.60 & 1.05 \\
    %NU-Wave & $3.00\times 10^6$ & 26.90 & 1.22 & 23.50 & 1.51 & 21.40 & 1.66 & 18.70 & 1.84 \\
    NU-Wave 2 & 1.70M & \textbf{28.40} & 0.77 & \textbf{24.00} & 0.93 & 21.60 & 1.01 & 18.80 & 1.14 \\
    UDM+ & - & - & 0.64 & - & 0.79 & - & 0.84 & - & - \\
    \textbf{Proposed} & 103M & 26.26 & \textbf{0.61} & 23.46 & \textbf{0.69} & \textbf{21.74} & \textbf{0.77} & \textbf{18.93} & \textbf{0.81} \\ \hline
    \end{tabular}
    }
    \label{tbl:res_A}
\end{minipage}
\hfill
\begin{minipage}{0.41\textwidth}
    \centering   \includegraphics[width=0.82\linewidth]{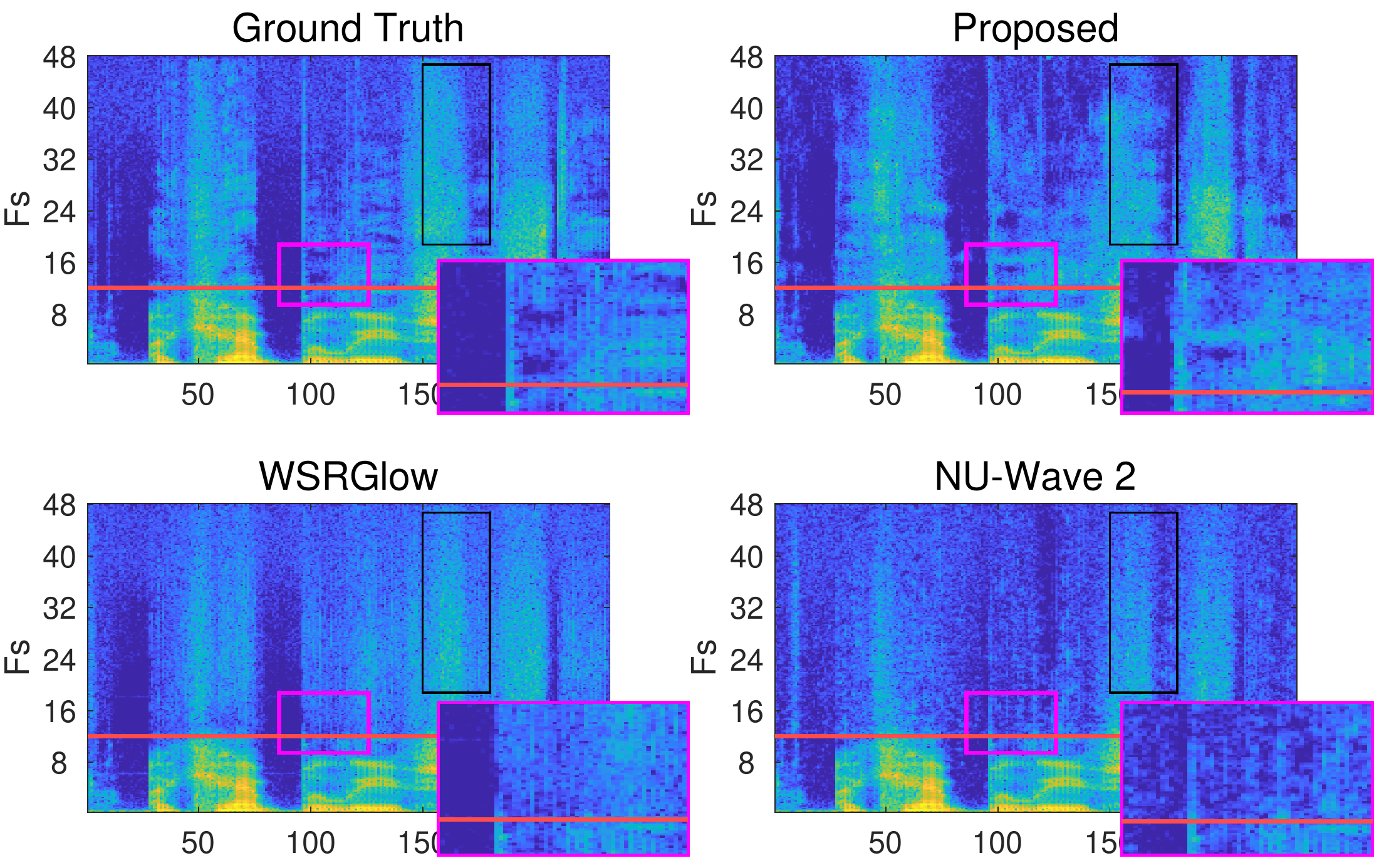}
    \captionof{figure}{Visualised comparison of $4\times$ SR. Note that our model produces richer harmonics (better zoom in).}
    \vspace{-0.4cm}
    \label{fig:sr_result1}
\end{minipage}
\end{center}

\begin{minipage}[t]{0.55\columnwidth}
\centering
\caption{MOS ($\uparrow$)}
\resizebox{!}{0.038\textheight}{
    \begin{tabular}{c|c|ccc}
    \hline
    Target & Input & LR & HR & SR \\ \hline\hline
    \multirow{4}{*}{48 kHz}& 8 kHz &  $2.5$ & $4.5$ & $3.8$ \\
    &12 kHz & $2.8$ & $4.8$ & $4.2$ \\
    &16 kHz & $3.0$ & $4.8$ & $4.5$ \\
    &24 kHz & $4.0$ & $4.8$ & $4.7$ \\
    \hline
    \end{tabular}
}\label{tbl:res_MOS}
\end{minipage}
\hfill
\begin{minipage}[t]{0.55\columnwidth}
\centering
\caption{PESQ-wb ($\uparrow$)}
    \resizebox{!}{0.038\textheight}{
    \begin{tabular}{c|c|c}
    \hline
    Target & Models & PESQ-wb\\ \hline\hline
    \multirow{4}{*}{\begin{tabular}[c]{@{}c@{}}16 kHz\\ ($2\times$ SR)\end{tabular}} & UDM+ &2.93 \\
    &NU-Wave 2 & 3.38 \\
    &NVSR & 3.47 \\
    &\textbf{Proposed} & \textbf{3.50} \\
    \hline
    \end{tabular}\label{tbl:res_PESQ}
}
\end{minipage}
\hfill
\begin{minipage}[t]{0.9\columnwidth}
\centering
\caption{Ablation studies for 8 kHz to 48 kHz SR.}
    \renewcommand{\arraystretch}{1.0}
    \resizebox{!}{0.038\textheight}{%
    \begin{tabular}{cc|cc}
    \hline
    Model & \# Parameters & $SNR\uparrow$ & $LSD\downarrow$ \\ \hline\hline
    \textbf{Proposed} & 103M & \textbf{18.92} & \textbf{0.81} \\
    w/o $\plog$ & 103M& \multicolumn{2}{c}{Failed to train}\\
    w/o pre-train & 103M & 11.25 (-7.67) & 0.84 (+0.03) \\
    w/o Transf blocks & 143M & 11.47 (-7.45) & 1.60 (+0.79) \\
    %w/o modified & \multirow{2}{*}{$7.23\times 10^7$}  %&\multirow{2}{*}{17.55 (-1.37)} & \multirow{2}{*}{0.88 (+0.07)}\\
    %up/down sampling & & &\\
    w/o Interp up-sampling & 72.3M & 17.55 (-1.37) & 0.88 (+0.07)\\
    \hline
    \end{tabular}%
    }
    \label{tbl:ablation}
\end{minipage}
\vspace{-0.2cm}
\end{table*}

\subsection{Dataset \& Pre-processing}

In this study, we train our model on a dataset composed of the CSTR's VCTK speech dataset~\cite{WaveNet} and the Hi-Fi TTS dataset~\cite{HiFi-TTS_data}. The sampling rate of this joint dataset is at least 44.1 kHz, and the total duration is up to 292 hours, ensuring a high-resolution and high-quality speech corpus.

We randomly select a 32512-point clip from each input $HR$ audio. We construct the $(\bm{LR},\bm{HR})$ training pair by filtering out signal components above the Nyquist frequency of $\bm{LR}$ to simulate the loss of high-frequency components during down-sampling. All filtering configurations use the default values of \texttt{torchaudio.functional.resample()}. For the MDCT layer, we set the frame length $N = 512$, which yields 256 points per frame after the FFT; the hop length $H = 256$, which produces 128 frames for a 32512-point segment. Thus, a single spectrogram has a size of $128\times 256$.

\subsection{Evaluation metrics}
Following previous works~\cite{AECNN-v,AFiLM,NU-Wave,nvsr}, we use the signal-to-noise ratio (SNR) and Log-spectral distance (LSD) as evaluation metrics to assess the proposed model.
% \begin{comment}
Specifically, given a reference signal $\mathbf{x}$ and a corresponding approximation $\mathbf{\hat{x}}$,  SNR is given by 
\small
\begin{equation}
%\operatorname{SNR}(x, y)=10 \log 
SNR(\mathbf{\hat{x}}, \mathbf{x})=10 \log_{10}
\frac{||\mathbf{x}||_{2}^{2}}{||\mathbf{x}-\mathbf{\hat{x}}||_{2}^{2}}
\end{equation}
\normalsize
%For the target signal $x$ and the model output estimate $\hat{x}$ , 
The LSD is defined as
\small
\begin{equation}
L S D(\mathbf{x}, \mathbf{\hat{x}})=\frac{1}{T} \sum_{t=1}^{T} \sqrt{\frac{1}{F} \sum_{f=1}^{F}  \left(\log_{10}\frac{\mathbf{X}^{2}[t, f]}{\hat{\mathbf{X}}^{2}[t, f]}\right)^{2}}
\end{equation}
\normalsize
where $T$ represents the period, $\mathbf{X}$ and $\hat{\mathbf{X}}$ represent the magnitude spectra of $\mathbf{x}$ and $\hat{\mathbf{x}}$, respectively, $t$ and $f$ are the index of frame and frequency, respectively. 
% \end{comment}
A lower LSD score and higher SNR value indicate a better SR performance. We use the averaged LSD and SNR scores of the VCTK-test as the final result of our model.

In addition to objective evaluation metrics, we also used subjective Mean Opinion Score (MOS) and Wide-band (8k$\rightarrow$16k) Perceptual Evaluation of
Speech Quality (PESQ-wb) to assess the quality of generated SR audio and compared it to that of the original HR audio. %, as shown in Table~\ref{tbl:res_B}.

%\subsection{Baselines}

\subsection{Training methods and techniques}
For our proposed model, we first pre-train it on a joint dataset of HiFi-TTS+VCTK with 120 epochs for learning SSR up to 44.1 kHz. After 60 epochs, the learning rate is linearly reduced to 0. We then fine-tune the model with 80 epochs to learn SSR up to 48 kHz by using only the VCTK part of pre-training dataset with 48 kHz audio only. During fine-tuning, the encoders and bottlenecks in $G_{global}$ and $G_{local}$ are frozen. After 40 epochs, the learning rate is linearly reduced to 0. All models were trained on an Nvidia RTX3090 GPU using an Adam optimiser~\cite{Adam} with $\beta_1=0.5$, $\beta_2=0.999$ and $\epsilon=10^{-6}$. And Automatic Mixed Precision (AMP)~\cite{AMP} is enabled.

\subsection{Results}

% \begin{table}
% %\begin{adjustbox}{width=0.8\columnwidth,center}
% %\caption{Subjective evaluation of the generated audio}
% %\label{tbl:res_B}
% \begin{minipage}[t]{0.49\columnwidth}
% \caption{Results of MOS ($\uparrow$) at 48 kHz target}
% \centering
% %\resizebox{!}{0.040\textheight}
% {%
%     \begin{tabular}{c|ccc}
%     \hline
%     Input & LR & HR & SR \\ \hline\hline
%     8k &  $2.5$ & $4.8$ & $3.8$ \\
%     12k & $2.8$ & $4.8$ & $4.2$ \\
%     16k & $3.0$ & $4.8$ & $4.5$ \\
%     24k & $4.0$ & $4.5$ & $4.7$ \\
%     \hline    \end{tabular}
% }\label{tbl:res_MOS}
% \end{minipage}
% \hfill
% \begin{minipage}[t]{0.45\columnwidth}
% \caption{PESQ-wb ($\uparrow$) scores for 8 kHz to 16 kHz}
% \centering
% %\resizebox{!}{0.040\textheight}
% {%
% \begin{tabular}{c|c}
% \hline
% Models & PESQ-wb \\ \hline\hline
% UDM+ & 2.93 \\
% NU-Wave 2 & 3.38 \\
% NVSR & 3.47 \\
% \textbf{Proposed} & \textbf{3.50} \\
% \hline
% \end{tabular}\label{tbl:res_PESQ}
% }
% \end{minipage}
% \vspace{-0.4cm}
% \end{table}
We chose several state-of-the-art methods as baseline models to compare with our mdctGAN, including AudioUnet~\cite{V.Kuleshov}, MUGAN~\cite{MUGAN}, %AECNN-varietas~\cite{AECNN-v},
WSRGlow~\cite{WSRGlow}, NU-Wave 2~\cite{NU-Wave-2}, NVSR~\cite{nvsr} %TFiLM~\cite{TFiLM}, 
% AFiLM~\cite{AFiLM}, 
 and UDM+\cite{udm+}. All models are using a 48 kHz output target. Here, the performance of these baseline models on the VCTK dataset~\cite{WaveNet} is measured using the published models and results of respective authors.
Table~\ref{tbl:res_A} compares the mdctGAN with other baseline models at a target of 48 kHz using SNR and LSD scores with different input sampling rates. Our model achieves the best LSD scores in all cases, especially for lower input sampling rates. Specifically, our model is more advantageous at 12 kHz and 8 kHz inputs that provides gains of 0.07dB and 0.33dB over the second-best models, respectively. In terms of SNR, our model also yields the best results for 12 kHz and 8 kHz inputs. Both WSRGlow and NU-Wave 2 exceed our proposed model at 24 kHz input and NU-Wave 2 is slightly better than ours at 16 kHz. Note that the mdctGAN is more memory efficient than WSRGlow in terms of the number of model parameters. Figure~\ref{fig:sr_result1} compares our output with others and the ground truth. And our model produces richer harmonics with greater high-frequency details than competitors. Table~\ref{tbl:res_MOS} shows that our model achieved high MOS that are consistent with the ground truth at various input sampling rates. For 16 to 48 kHz SR, our method are also competitive compared to recent works\cite{Nonparallel}, with Hifi-GAN, WSRGlow, and Dual-Cycle-GAN achieving mean MOS of 4.23, 4.23, and 4.51, respectively. Moreover, our method outperformed competing methods in terms of PESQ scores, shown in Table~\ref{tbl:res_PESQ}. This indicates that our model is able to generating SR audio that is close to the natural listening quality. %the quality of SR is also very close to that of HR by the scores of the subjective metric MOS.

% \begin{table}[!htb]
% \centering
% \caption{SNR/LSD results of ablation studies for 48 kHz target and 8 kHz input.}
% % \vspace{-0.1cm}
% \renewcommand{\arraystretch}{1.0}
% \resizebox{\columnwidth}{!}{%
% \begin{tabular}{cc|cc}
% \hline
% Model & \# Parameters & $SNR\uparrow$ & $LSD\downarrow$ \\ \hline\hline
% \textbf{Proposed} & 103M & \textbf{18.92} & \textbf{0.81} \\
% w/o $\plog$ & 103M& \multicolumn{2}{c}{Failed to train}\\
% w/o pre-train & 103M & 11.25 (-7.67) & 0.84 (+0.03) \\
% w/o Transf blocks & 143M & 11.47 (-7.45) & 1.60 (+0.79) \\
% %w/o modified & \multirow{2}{*}{$7.23\times 10^7$}  %&\multirow{2}{*}{17.55 (-1.37)} & \multirow{2}{*}{0.88 (+0.07)}\\
% %up/down sampling & & &\\
% w/o Interp up-sampling & 72.3M & 17.55 (-1.37) & 0.88 (+0.07)\\
% \hline
% \end{tabular}%
% }
% \label{tbl:ablation}
% \vspace{-0.4cm}
% \end{table}
\subsection{Ablation Study}\label{sec:ablation}
To verify the effectiveness of the key components in our model, different configurations are evaluated as follows: \emph{i}) removing the $\plog$ dynamic compressing ; \emph{ii}) using only VCTK dataset for network training; \emph{iii}) substituting all Transformer Blocks with ResNet Blocks; \emph{iv}) using transpose-conv for up-sampling. Table~\ref{tbl:ablation} indicates that without $\plog$ compression, the network cannot be trained. Pre-training on a larger dataset substantially increases the network's overall performance. In addition, the addition of Transformer Blocks results in considerable improvements in both SNR and LSD values. Furthermore, with interpolation up-sampling, better performance are obtained at the expense of the increased number of parameters.

\section{Conclusion and Future work}
We present \textbf{mdctGAN}, a novel SSR method adapting a transformer-based GAN to reconstruct high-quality speech. It works on MDCT domain without additional phase estimation to recover raw waveforms. %We also introduce pseudo-log dynamic range compression for a better representation of the patterns in the MDCT spectrograms. 
By incorporating MDCT with multiple critical enhancements, including pseudo-log compression and Transformer blocks, we have successfully proposed an SSR framework and evaluated it on the VCTK test dataset. mdctGAN outperformed previous models for 48 kHz target with various input resolution settings and achieved state-of-the-art LSD scores. The quality of our model's results was further validated by subjective metrics, MOS and PESQ.

Despite the many advantages of our proposed approach, there is still room for improvement. Our model needs to be trimmed for real-time processing. Moreover, the SNR is not optimal at low input sampling rates. In the future, we plan to improve mdctGAN to make it more compact and enhance its SR quality. We also encourage further research to follow our proposed MDCT-based approach to achieve better speech enhancement.

\bibliographystyle{IEEEtran}
\bibliography{mybib}

\end{document}